\documentclass[aps,epsf,twocolumn,prl,superscriptaddress,showpacs]{revtex4}
\usepackage{graphicx} 
\usepackage{dcolumn} 
\usepackage{bm} 
\usepackage{longtable} 

\begin{document}

\tighten                                                 

\draft
%


%
\title
{
Compton Scattering from the Deuteron and Extracted Neutron Polarizabilities 
}

\author
{M. Lundin}
\affiliation
{Department of Physics, University of Lund, SE-22100 Lund, Sweden}
\author
{J.-O. Adler}
\affiliation
{Department of Physics, University of Lund, SE-22100 Lund, Sweden}
\author
{M. Boland}
\affiliation
{Department of Physics, University of Lund, SE-22100 Lund, Sweden}
\author
{K. Fissum}
\affiliation
{Department of Physics, University of Lund, SE-22100 Lund, Sweden}
\author
{T. Glebe}
\affiliation
{Zweites Physikalisches Institut, Universit\"at G\"ottingen 
D-37073 G\"ottingen, Germany}
\author
{K. Hansen}
\affiliation
{Department of Physics, University of Lund, SE-22100 Lund, Sweden}
\author 
{L.  Isaksson}
\affiliation
{Department of Physics, University of Lund, SE-22100 Lund, Sweden}
\author
{O.~Kaltschmidt}
\affiliation
{Zweites Physikalisches Institut, Universit\"at G\"ottingen 
D-37073 G\"ottingen, Germany}
\author
{M. Karlsson}
\affiliation
{Department of Physics, University of Lund, SE-22100 Lund, Sweden}
\author
{K.~Kossert}
\affiliation
{Zweites Physikalisches Institut, Universit\"at G\"ottingen 
D-37073 G\"ottingen, Germany}
\author
{M.I. Levchuk}
\affiliation
{B.I. Stepanov Institute of Physics, Belarussian Academy of
Sciences, 220072 Minsk, Belarus}
\author
{P. Lilja}
\affiliation
{Department of Physics, University of Lund, SE-22100 Lund, Sweden}
\author
{B. Lindner}
\affiliation
{Kristianstad University College, SE-29188 Kristianstad, Sweden}
\author
{A.I.  L'vov}
\affiliation
{P.N. Lebedev Physical Institute, 119991 Moscow, Russia}
\author 
{B. Nilsson}
\affiliation
{Department of Physics, University of Lund, SE-22100 Lund, Sweden}
\author
{D.E. Oner}
\affiliation
{Zweites Physikalisches Institut, Universit\"at G\"ottingen 
D-37073 G\"ottingen, Germany}
\author
{C. Poech}
\affiliation
{Zweites Physikalisches Institut, Universit\"at G\"ottingen 
D-37073 G\"ottingen, Germany}
\author
{S. Proff}
\affiliation
{Zweites Physikalisches Institut, Universit\"at G\"ottingen 
D-37073 G\"ottingen, Germany}
\author
{A.~Sandell}
\affiliation
{Department of Physics, University of Lund, SE-22100 Lund, Sweden}
\author
{B.  Schr\"oder}
\affiliation
{Department of Physics, University of Lund, SE-22100 Lund, Sweden}
\author
{M. Schumacher}
\affiliation
{Zweites Physikalisches Institut, Universit\"at G\"ottingen 
D-37073 G\"ottingen, Germany}
\author
{D.A. Sims}
\affiliation
{Department of Physics, University of Lund, SE-22100 Lund, Sweden}

\date{\today}
%
\begin{abstract}

Differential cross sections for Compton scattering from the deuteron
were measured at MAX-lab for incident photon energies of 55 MeV and 66 MeV
at nominal laboratory angles  of $45^\circ$, $125^\circ$, and $135^\circ$. 
Tagged photons were scattered 
from liquid deuterium and detected in three NaI spectrometers.
By comparing the data with theoretical calculations in the framework of 
a one-boson-exchange potential model, the sum and difference of the 
isospin-averaged nucleon polarizabilities,  
$\alpha_N + \beta_N = 17.4 \pm 3.7$ and 
$\alpha_N - \beta_N =  6.4 \pm 2.4$ (in 
units of $10^{-4}$ fm$^3$), have been determined. By combining the latter 
with the global-averaged value for $\alpha_p - \beta_p$
and using the predictions of the Baldin sum rule
for the sum of the nucleon polarizabilities,
we have obtained values for 
the neutron electric and magnetic polarizabilities of
$\alpha_n=~8.8~\pm~2.4$(total) $\pm~3.0$(model) and 
$\beta_n =~6.5~ \mp~2.4$(total) $\mp~3.0$(model), respectively.

\end{abstract}


\maketitle

The electric ($\alpha$) and magnetic ($\beta$) polarizabilities of the nucleon 
characterize the second-order response of its internal 
structure to applied electric and magnetic fields, respectively.  
Since the polarizabilities manifest themselves in two-photon 
processes, an excellent method to measure them is via Compton 
scattering experiments.

The most recent global average~\cite{Olmos01} for the 
difference of the proton polarizabilities is
\begin{equation}
\alpha_p-\beta_p=
10.5\pm 0.9 ({\rm stat+syst})\pm 0.7 ({\rm model}),
\label{olmos} 
\end{equation}
in units of $10^{-4}$ fm$^3$ (which will be used throughout this
paper). 
The sum of the nucleon polarizabilities is
usually obtained indirectly via the predictions of the
Baldin sum rule.  A recent re-evaluation of this 
sum rule~\cite{LL00} gives
\begin{eqnarray}
\alpha_p +\beta_p &=&14.0\pm 0.3, \label{apbp}\\
\alpha_n +\beta_n &=&15.2\pm 0.5. \label{apbn}
\end{eqnarray}

Electromagnetic scattering of low-energy neutrons in the electric 
fields of heavy nuclei was used to extract $\alpha_n$ (see Ref.~\cite{aleks66} and the references therein).  
There have been many attempts to measure $\alpha_n$ 
using this method \cite{aleksandr,Sch91,Koe95,Eni97}, 
but the final conclusion regarding its value 
has remained  unclear.  Note that this method does not constrain 
$\beta_n$.

Scattering real photons from neutrons bound in nuclei is
another way to measure the neutron polarizabilities.
The use of a deuterium target permits the theoretical uncertainties
in the interpretation of the experimental data to be minimized.
Both the reactions $\gamma d\to \gamma 
np$ (quasi-free) and $\gamma d\to \gamma d$ (elastic) can be considered.
The suggestion to exploit the quasi-free 
kinematic region was made in Refs.~\cite{levchuk94, wissmann98}.
An advantage of this approach is that 
it can be carefully tested by comparing measured
quasi-free proton cross sections with available free-proton data. The method 
has been used in a series of experiments 
at MAMI~\cite{rose90,wissmann99,kossert02} and SAL~\cite{kolb00}.  The values 
for the neutron polarizabilities found in these experiments are in 
agreement with each other; however, the sizes of the 
quoted errors differ markedly.  The most accurate values to date 
have recently been reported in Ref.~\cite{kossert02} as
\begin{eqnarray}
\alpha_n&=&12.5\pm 1.8 ({\rm stat}){}^{+1.1}_{-0.6}{} ({\rm syst})
\pm 1.1 ({\rm model}), \label{aln} \\ 
\beta_n &=&~\,2.7\mp 1.8 ({\rm stat}){}^{+0.6}_{-1.1}{} ({\rm syst})
\mp 1.1 ({\rm model}). \label{btn}
\end{eqnarray}
The anti-correlated errors in Eqs.~(\ref{aln}) and  (\ref{btn}) are 
due to the application of the sum-rule result (\ref{apbn}).
In view of the model errors contained  in 
Eqs.~(\ref{aln}) and (\ref{btn}), 
confirmation of these values is of great importance. 

Elastic photon scattering from the deuteron provides a
third experimental method for determining the neutron polarizabilities. While only
the isospin-averaged nucleon polarizabilities
$\alpha_N=(\alpha_p+\alpha_n)/2$ and $\beta_N=(\beta_p+\beta_n)/2$
can be measured, this is not a major problem
since the proton values are rather accurate (see Eqs.~(\ref{olmos})
and (\ref{apbp})). Although the first such measurements of elastic $\gamma d$ scattering
were performed many years
ago (see Ref.~\cite{tenore65} and the references therein), only two recent
experiments at Illinois at E$_\gamma$ = 49 and 69 MeV~\cite{Lucas94}
and at SAL at E$_\gamma$ = 94 MeV~\cite{Horn00} have been precise enough to reveal the
effect of the nucleon polarizabilities.  However, the values for
$\alpha_N-\beta_N$ extracted from these measurements
are inconsistent. With the use of the theoretical model~\cite{LL00},
a value of $2.6\pm 1.8$ was obtained in
Ref.~\cite{Horn00} that together with Eq.~(\ref{olmos}) gives
$\alpha_n-\beta_n\simeq - 5.3\pm 3.8$. But to describe the data from
Ref.~\cite{Lucas94}, one needs to increment $\alpha_N-\beta_N$ 
to $7.9\pm 3.8$, thus giving $\alpha_n-\beta_n = 5.3\pm 7.6$.  
The former value is far from theoretical estimates of this quantity based on dispersion
relations, which crudely predict that $\alpha_n-\beta_n \simeq
\alpha_p-\beta_p$ (see Refs.~\cite{Lvov93,petr81,HN94}).
It also contradicts the results obtained from
quasi-free neutron Compton scattering (see Eqs.~(\ref{aln}) and  (\ref{btn})).

In this Letter, we report a new measurement of the differential
cross section for deuteron Compton scattering performed at
MAX-lab. The near-continuous, 95 MeV electron beam from the MAX I stretcher
ring was used to produce tagged photons, in the energy range 50-72
MeV, with an FWHM energy resolution of 
$\sim$330 keV and a flux of about 
$3\times 10^5$ MeV$^{-1}$s$^{-1}$~\cite{Lin90,Adl97}.  The post-bremsstrahlung electrons were
momentum analysed in a magnetic spectrometer using two 32 scintillator hodoscopes
located along the focal plane. They were placed such that the central
tagged-photon energies were 55 MeV and 66 MeV. The photon beam was
incident upon a
scattering chamber containing liquid deuterium in a cylindrical cell (length 160 mm and diameter 48 mm)~\cite{Gle93}
made from 125 $\mu$m thick Kapton foil.

Scattered photons were detected in three spectrometers, each
containing a central NaI detector 25.4 cm in diameter and with depths
of either 25.4 or 35.5 cm, placed at nominal lab angles of 45$^\circ$, 125$^\circ$, and
135$^\circ$ at a distance of approximately 0.4 m from the target.
The resulting solid angle (together with the detection efficiency)
was precisely determined via GEANT simulations. The energy
resolution of the NaI spectrometers ranged from 6 to 8\% at a photon
energy of 60 MeV. The gain stability of the NaI detectors was continuously monitored
and instabilities were corrected for using a Light Emitting Diode (LED) system
\cite{Lun02}. In turn, the stability of the LED system was monitored and verified by
examining the location of the cosmic peak on a run-by-run basis.  
The data acquisition was started by an event in any one of
the NaI detectors, which provided gates for charge integrating ADCs 
(Analog to Digital Converter) and
start signals for TDCs (Time to Digital Converter) used for time-of-flight (TOF) measurements. The
stop signals for the TDCs came from the focal-plane detectors. 
In order to monitor the number of pile-up events in
the NaI crystals, a 250 MHz Flash ADC was used. Less than 1\% 
of the events were affected by pile-up. 

\begin{figure} \resizebox{0.45\textwidth}{!}{\includegraphics{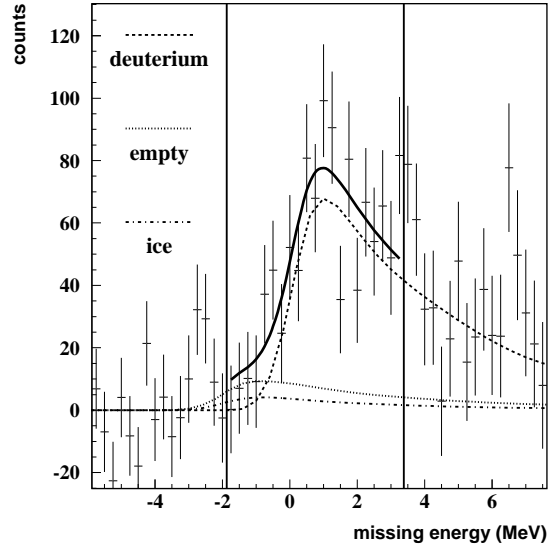}}
\caption{\label{fig:spectrum}$E_{\rm{miss}}$ spectrum at $\theta_{\gamma}^{\rm{lab}}=126^\circ$ and $E_\gamma=55$ MeV.
The two vertical lines indicate the region of interest (ROI) used to determine 
the yield. The solid line represents the sum of the fitted responses (see text).
 } \end{figure}

Data were collected over 8 weeks, divided typically into two-week run periods. 
For each run period, tagging
efficiencies ($\sim$20\%) were measured using a Pb/SCIFI
detector~\cite{Her90}. In addition, the responses of the NaI spectrometers
were measured by placing them directly in the reduced intensity photon beam. For
each detector, the aforementioned GEANT simulations were phenomenologically broadened
to match the measured responses. This broadening was then folded
into second-stage simulations of the in situ
responses and solid angles
~\cite{Lun02}. A typical result is shown by the dashed line in Fig.~\ref{fig:spectrum}.

This missing energy ($E_{\rm{miss}}$) spectrum was obtained by
summing over a 10 MeV interval 
centered at 55 MeV.  $E_{\rm{miss}}$ was defined as the difference between
the tagged-photon energy (corrected for the Compton scattering energy
shift) and the energy registered in the NaI detector.
The subtraction of random events was performed using two
independent methods: one used the TOF between the
scattered photons and the post-bremsstrahlung electrons, while the other
employed the non-physical ($E_{\rm{miss}} <$ 0) region of the energy spectrum. 
An average normalization factor was employed which resulted
in a 4\% systematic uncertainty in the extracted cross sections. The
experimental data were fitted within the region of interest (ROI) considering the contributions
of elastically scattered photons from liquid deuterium, from the
Kapton cell and the scattering chamber windows, 
and from ice (H$_2$O) which built up on the cell during the
run period. The contribution from the cell and windows was
calculated using known differential cross sections for carbon and
oxygen~\cite{Lud91,Hag95,Pro98}. 
The contribution from ice was fitted using data from the largest detector
at the backward angles and extrapolated to the other detector angles. The careful quantification of the
experimental background due to the empty target and ice resulted in an average
improvement in the $\chi^{2}$ of the fit functions of 20\% for the backward 
detector angles. The observed contributions
correspond to ice-layer thicknesses of about 100 $\mu$m per cell endcap.
The contribution
from inelastic scattering (which begins at $E_{\rm{miss}}$ = 2.2 MeV)
was kept to less than 3\% via the narrow  ROI.
\begingroup
\squeezetable
\begin{table}[hbt]
\begin{ruledtabular}
\caption{CM differential cross sections for deuteron
Compton scattering.
The first uncertainty is statistical and the second is systematic.}
\label{crsect}
\begin{tabular}{rrrc}
$E_{\gamma}^{\rm{lab}}$(MeV)&$\Theta_{\gamma}^{\rm{lab}}$(deg) 
& $\Theta_{\gamma}^{\rm{CM}}$(deg) & 
d$\sigma^{\rm{CM}}$/d$\Omega_\gamma$ (nb/sr)\\
\hline
54.6   &  43.8   &  44.9   &  16.8 $\pm$ 4.1 $\pm$ 1.5 \\
54.6   & 123.7   & 125.0   &  15.7 $\pm$ 1.5 $\pm$ 1.3 \\
54.6   & 135.7   & 136.8   &  17.2 $\pm$ 2.0 $\pm$ 1.4 \\
54.9   &  43.2   &  44.3   &  16.6 $\pm$ 3.3 $\pm$ 1.8 \\
54.9   & 126.3   & 127.6   &  15.4 $\pm$ 1.3 $\pm$ 1.0 \\
54.9   & 135.2   & 136.3   &  18.4 $\pm$ 1.7 $\pm$ 1.6 \\
55.9   &  48.9   &  50.2   &  13.4 $\pm$ 2.7 $\pm$ 1.0 \\
55.9   & 130.4   & 131.7   &  15.3 $\pm$ 2.0 $\pm$ 1.2 \\
55.9   & 136.2   & 137.3   &  21.0 $\pm$ 3.2 $\pm$ 2.2 \\
65.3   &  43.6   &  44.9   &  16.0 $\pm$ 2.8 $\pm$ 1.4 \\
65.3   & 123.7   & 125.3   &  15.3 $\pm$ 1.3 $\pm$ 1.4 \\
65.3   & 135.5   & 136.8   &  12.6 $\pm$ 1.7 $\pm$ 1.8 \\
65.6   &  43.1   &  44.4   &  18.6 $\pm$ 2.4 $\pm$ 1.4 \\
65.6   & 126.3   & 127.8   &  16.0 $\pm$ 1.2 $\pm$ 1.1 \\
65.6   & 135.2   & 136.5   &  15.1 $\pm$ 1.7 $\pm$ 1.3 \\
67.0   &  48.8   &  50.3   &  15.2 $\pm$ 1.8 $\pm$ 1.2 \\
67.0   & 130.3   & 131.8   &  14.2 $\pm$ 1.5 $\pm$ 1.0 \\
67.0   & 136.1   & 137.5   &  15.0 $\pm$ 2.7 $\pm$ 1.2 \\
\end{tabular}
\end{ruledtabular}
\end{table}
\endgroup
The systematic uncertainties for the experiment ($\sim$8\%) arise 
from the tagging efficiency  (5\%), the
product of the solid angle and the detection efficiency (4\%), 
the random background subtraction (4\%), the contamination of the ROI by
 inelastic photon scattering (3\%), and the target thickness
 (2\%). 

\begin{figure} \resizebox{0.40\textwidth}{!}{\includegraphics{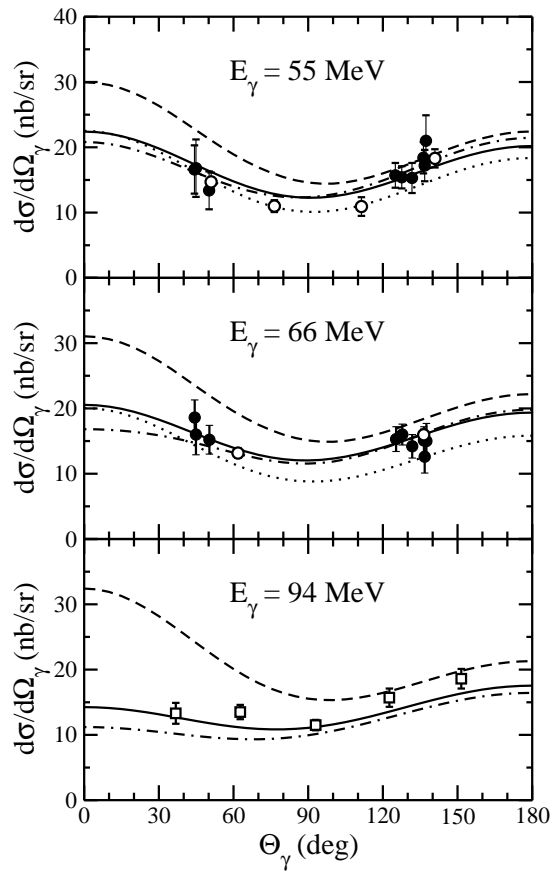}} 
\caption
{\label{fig:results}CM differential cross sections for Compton 
scattering from the deuteron.  
Filled circles: present experiment; 
open circles: extrapolated 
Illinois results~{\protect \cite{Lucas94}} (see text); open squares: SAL results {\protect \cite{Horn00}}. 
The error bars represent the quadratic sum of the statistical and systematic uncertainties.
Solid and dashed curves are the calculations of Ref.~{\protect \cite{LL00}}
(dashed: $\alpha_N$ = $\beta_N$ = 0; solid $\alpha_N$ = 11.9, $\beta_N$ = 5.5, see
Eqs.~(\ref{apbN}), (\ref{ambN})).  Dotted and dashed-dotted curves are the 
predictions of Refs.~\cite{grie02} and \cite{bean02}, respectively, extrapolated
to $\alpha_N$ = 11.9, $\beta_N$ = 5.5.
} 
\end{figure}

Results for the center-of-mass (CM) cross sections are given in
Table~\ref{crsect} and displayed in Fig.~\ref{fig:results}.
Earlier results from Illinois~\cite{Lucas94} and SAL~\cite{Horn00} are also shown.
Since the Illinois data were obtained at slightly different photon energies (49 MeV and
69 MeV), we 
extrapolated them to our energies using the theoretical model presented in 
Ref.~\cite{LL00}.
Good agreement with the Illinois data is clearly demonstrated.

The polarizabilities of the nucleon have been determined using the 
theoretical model presented in Ref.~\cite{LL00}. In this model, apart from
effects due to the one-body $\gamma$$N$ interaction (which contain the 
effects due to the polarizabilities), two-body effects due to the $NN$ interaction
and related Meson-Exchange Currents (MEC) have been included via 
a series of one-boson exchanges which constitute the Bonn potential. 
After making a two-parameter fit to the data in Table~\ref{crsect} with the
use of this model, we obtain for the sum and difference of the isospin-averaged
nucleon polarizabilities
\begin{eqnarray}
\alpha_N + \beta_N &=&  17.4 \pm 3.7 ({\rm stat+syst}), 
\label{apbN}\\
\alpha_N - \beta_N &=&~\,6.4 \pm 2.4 ({\rm stat+syst}), 
\label{ambN}
\end{eqnarray}
with $\chi^2/N_{\rm{dof}}=7.5/(18-2)$.
The quoted uncertainties are the statistical and systematic uncertainties
taken in quadrature.
The obtained sum $\alpha_N + \beta_N$ is in agreement with 
the values given in Eqs.~(\ref{apbp}) and (\ref{apbn}).
This indicates that the systematic uncertainties
(including the model uncertainty) are well understood.

Having extracted $\alpha_N$ and $\beta_N$, we invoked 
the accurate proton values for $\alpha_p$ and $\beta_p$ 
to derive  the neutron polarizabilities. In doing this,
we relied on the precise sum-rule prediction  
(\ref{apbn}) rather than on our result (\ref{apbN}).
 Making use of 
Eqs.~(\ref{olmos})--(\ref{apbn}) and (\ref{ambN}), we obtain
\begin{eqnarray}
\alpha_n &=& 8.8 \pm 2.4({\rm stat+syst}), \label{an}\\ 
\beta_n &=& 6.5 \mp 2.4({\rm stat+syst}). \label{abn}
\end{eqnarray}

If all currently available data (Illinois, SAL, and Lund) are fit using the 
theoretical model of Ref.~\cite{LL00}, the following ``global values" may be inferred:
\begin{eqnarray}
\alpha_N + \beta_N &=&  16.7 \pm 1.6 ({\rm tot}), 
\label{aNpbN}\\
\alpha_N - \beta_N &=&~\,4.8 \pm 2.0 ({\rm tot}), 
\label{aNmbN}
\end{eqnarray}
with $\chi^2/N_{\rm{dof}}=38/(29-2)$. Thus, using the Baldin sum rule and Eq. (\ref{olmos}),
\begin{eqnarray}
\alpha_n  &=&  7.2 \pm 2.1 ({\rm tot}), 
\label{anfinal}\\
\beta_n &=&  8.1 \mp 2.1 ({\rm tot}). 
\label{bnfinal}
\end{eqnarray}
Here both statistical and systematic uncertainties have been combined, the 
latter being taken into account through a rescaling of measured
cross sections within their normalization uncertainties.

Model uncertainties in the extracted values of the polarizabilities can be partly understood by
comparing different calculations of 
d$\sigma$/d$\Omega$ \cite{LL00,chen98,kara99,bean99,grie02,bean02}.
Fig.~\ref{fig:results} shows the three most recent predictions  \cite{LL00,grie02,bean02} at 
fixed $\alpha_N$ = 11.9 and $\beta_N$ = 5.5 as given by 
(\ref{apbN}) and (\ref{ambN}). At the energies and angles of the present experiment,
there is reasonable agreement between the potential model of Ref.~\cite{LL00} and the ${\cal O}(p^4)$
Chiral Perturbation Theory (ChPT) results of  Ref.~\cite{bean02}. 
At the SAL energy and angles, this agreement is poorer.
Agreement with the ${\cal O}(p^3)$ ChPT calculation of Ref.~\cite{grie02} is worse. 

Within the framework of the potential model used here, the main uncertainties which
contribute to the extraction of the polarizabilities arise from evaluations of the MEC and
seagull terms. The majority of these uncertainties may be collected into the 
Thomas-Reiche-Kuhn sum rule enhancement parameter $\kappa$ which determines
the magnitude of the seagull contribution \cite{LL00}. Depending on the $NN$ potential
used (Paris, Bonn, OBEPR, Argonne v18, Nijm93, or CD-Bonn) the value of $\kappa$ varies
from 0.44 to 0.51. This leads to variations in the differential cross section of about 4\%  which
result in variations in the extracted value of $\alpha_N$ of about $\pm$1.5.
We adopt this as an estimate of the model uncertainty in $\alpha_N$.  Accordingly,
the model uncertainty in the derived value of $\alpha_n$ is about $\pm$3.
The model uncertainty in $\beta_n$ is anticipated to be smaller than that
in $\alpha_n$, as the changes in $\kappa$ given above affect $\beta_N$
to a smaller degree.

The values obtained for $\alpha_n$ and $\beta_n$ are in reasonable agreement with those found in
quasi-free Compton scattering from the neutron (see Eqs.~(\ref{aln}) and  (\ref{btn})). However, we
observed the tendency of the technique of elastic Compton scattering from the deuteron
to give smaller values for $\alpha_n - \beta_n$ than did the technique of quasi-free scattering.
Predictions for $\alpha_n$ in the framework of ChPT at  ${\cal O}(p^4)$ \cite{BKM93}, the heavy
baryon ChPT model with the $\Delta$-isobar included \cite{hemmert98}, and the so-called ``covariant
dressed K-matrix model" \cite{kondr01} give values of 13.0 $\pm$ 1.5, 16.4, and 12.7,
respectively. All are somewhat larger than our value. The situation concerning $\beta_n$ is less
clear. Resonable agreement with ChPT results of 7.8 $\pm$ 3.6 \cite{BKM93}
and 9.1 \cite{hemmert98} is demonstrated. Ref.~\cite{kondr01} suggests a noticably smaller
value of 1.8.

In summary, differential cross sections 
for deuteron Compton scattering have been measured at MAX-lab.
The data were 
used to extract neutron polarizabilities in a model-dependent analysis. The extracted values 
for $\alpha_n$ and $\beta_n$ are consistent with those obtained from
the Illinois data~\cite{Lucas94}; however, they are inconsistent with those resulting
from an analysis of the higher energy SAL data \cite{Horn00}. The present analysis thus 
confirms the previous observations \cite{LL00,bean99} that the available models cannot
reconcile data obtained at photon energies of about 60 MeV with those
obtained at about 100 MeV.
   

{\small 
This project was supported by 
the Swedish Research Council, 
the Knut and Alice Wallenberg Foundation, 
the Crafoord Foundation, 
the Swedish Institute, the Wenner-Gren Foundation, and 
the Royal Swedish Academy of Sciences.
The participation
of the G\"ottingen group in the experiment was supported by
Deutsche Forschungsgemeinschaft.
}



\vspace{-5mm}
\begin{references}

\bibitem{Olmos01} V. Olmos de Le\'on {\it et al.}, Eur. Phys. J. A {\bf 10},
  207 (2001). 

\bibitem{LL00} M.I. Levchuk and A.I. L'vov, Nucl. Phys. {\bf A674}, 449 (2000);
M.I. Levchuk and A.I. L'vov, Nucl. Phys.  {\bf A684}, 490 (2001).

\bibitem{aleksandr} Yu.A. Aleksandrov, Phys. Part. Nucl. {\bf 32}, 708 (2001).

\bibitem{aleks66} Yu.A. Aleksandrov {\it et al.},
 JETP Lett. {\bf 4}, 134 (1966). 

\bibitem{Sch91} J. Schmiedmayer {\it et al.},
\prl {\bf 66}, 1015 (1991).
\bibitem{Koe95} L. Koester {\it et al.}, \prc {\bf 51}, 3363 (1995).
\bibitem{Eni97} T.L. Enik {\it et al.}, Sov. J. Nucl. Phys. {\bf 60}, 567 (1997).

\bibitem{levchuk94} M.I. Levchuk, A.I. L'vov, and V.A. Petrun'kin, 
         FIAN report No. 86, 1986;
         Few-Body Syst. {\bf 16}, 101 (1994).
\bibitem{wissmann98} F. Wissmann, M.I. Levchuk, and M. Schumacher, 
                 Eur. Phys. J. A {\bf 1}, 193 (1998).
\bibitem{rose90} K.W. Rose {\it et al.}, Phys. Lett. B {\bf 234}, 460 (1990);
                                         Nucl. Phys. {\bf A514}, 621 (1990).
\bibitem{wissmann99}
      F. Wissmann {\it et al.}, Nucl. Phys. {\bf A660}, 232 (1999).
\bibitem{kossert02} K. Kossert {\it et al.}, \prl {\bf 88}, 162301 
        (2002). 
\bibitem{kolb00} N.R. Kolb {\it et al.}, \prl {\bf 85}, 1388 (2000).


\bibitem{tenore65} A. Tenore and A. Verganelakis, Nuovo Cimento {\bf 
         35}, 261 (1965).
\bibitem{Lucas94} M.A. Lucas, Ph.D.\ thesis, University of Illinois, 1994.
\bibitem{Horn00} D.L. Hornidge {\it et al.}, \prl {\bf 84}, 2334 (2000).



\bibitem{petr81} V.A. Petrun'kin, Sov. J. Part. Nucl. {\bf 12}, 278 
         (1981).

\bibitem{Lvov93} A.I. L'vov, Int. J. Mod. Phys. A {\bf8}, 5267 (1993).
\bibitem{HN94} B.R. Holstein and A.M. Nathan, \prd {\bf 49}, 6101 (1994).


\bibitem{Lin90} L.J. Lindgren {\it et al.}, Nucl. Instrum. Methods Phys. Res. Sect. 
 A {\bf 294}, 10 (1990).
\bibitem{Adl97} J.-O. Adler {\it et al.}, Nucl. Instrum. Methods Phys. Res. Sect. 
 A {\bf 388}, 17 (1997).
\bibitem{Gle93} T. Glebe, MSc thesis, G\"ottingen University, 1993.

\bibitem{Lun02} M. Lundin, Ph.D. thesis, Lund University, 2002.

\bibitem{Her90} D.W. Hertzog {\it et al.}, Nucl. Instrum. Methods Phys. Res. Sect. 
 A {\bf 294}, 446 (1990).
\bibitem{Lud91} M. Ludwig {\it et al.}, Phys. Lett. B {\bf 274}, 275 (1992).
\bibitem{Hag95} D. H\"ager {\it et al.}, Nucl. Phys. {\bf A595}, 287 (1995).

\bibitem{Pro98} S. Proff {\it et al.}, Nucl. Phys. {\bf A646}, 67 (1999).

\bibitem{grie02} 
H.W. Griesshammer and G. Rupak, Phys. Lett. B {\bf 529}, 57 (2002).
\bibitem{bean02}
S.R. Beane {\it et al.}, arXiv:nucl-th/0209002.


\bibitem{chen98}
  J.-W. Chen {\it et al.}, Nucl. Phys. {\bf A644}, 245 (1998).

\bibitem{kara99}
  J.J. Karakowski and G.A. Miller,  \prc {\bf 60}, 014001 (1999).

\bibitem{bean99}
  S.R. Beane {\it et al.}, Nucl. Phys. {\bf A656}, 367 (1999);
M. Malheiro {\it et al.}, arXiv:nucl-th/0111047.


\bibitem{BKM93} 
  V. Bernard, N. Kaiser, and U.-G. Mei\ss ner, 
   Phys. Lett. B {\bf 319}, 269 (1993).
\bibitem{hemmert98}
   T.R. Hemmert {\it et al.}, \prd {\bf 57}, 5746 (1998).
\bibitem{kondr01}
   S. Kondratyuk and O. Scholten, \prc {\bf 64}, 024005 (2001).


\end{references}
\end{document}